\documentclass{elsart}
\usepackage{graphicx}

\begin{document}

\begin{frontmatter}

\title{Phantom shell around black hole and global geometry}

\author{V.~Berezin}, \ead{berezin@ms2.inr.ac.ru}
\author{V.~Dokuchaev}, \ead{dokuchaev@inr.npd.ac.ru}
\author{Yu.~Eroshenko} \ead{erosh@inr.npd.ac.ru} and
\author{A.~Smirnov} \ead{smirnov@ms2.inr.ac.ru}

\address{Institute for Nuclear Research of the Russian Academy
of Sciences, Moscow}

\begin{abstract}
We describe the possible scenarios for the evolution of a thin
spherically symmetric self-gravitating phantom shell around the
Schwarzschild black hole. The general equations describing the
motion of the shell with a general form of equation of state are
derived and analyzed. The different types of space-time $R_{\pm}$-
and $T_{\pm}$-regions and shell motion are classified depending on
the parameters of the problem. It is shown that in the case of a
positive shell mass there exist three scenarios for the shell
evolution with an infinite motion and two distinctive types of
collapse. Analogous scenarios were classified for the case of a
negative  shell mass. In particular this classification shows that
it is impossible for the physical observer to detect the fantom
energy flow. We shortly discuss the importance of our results for
astrophysical applications.
\end{abstract}


\end{frontmatter}


\section{Introduction}
Recent observations of both the type Ia high-redshift Supernovas (SNs) and Cosmic Microwave
Background (CMB) anisotropy at the small angular scales strongly indicate in favor of the
acceleration of the universe expansion at the present epoch \cite{acceler}. The simplest possibility for the acceleration expansion of the universe is an existence of the cosmological constant $\Lambda$ with an equation of state $w\equiv p/\rho=-1$ \cite{MisThoWhe}. The cosmological constant provides the satisfactory explanation of the cosmic dynamics but encounters the fine tuning problem.  An alternative explanation is the existence of dark energy in the form of a specific scalar field (quintessence) whose equation of state is varying with time (see e.~g. \cite{CaDaSt,ArMuSt}).  Such a model allows to construct the so called `tracker' or `attractor' cosmological solutions which resolve in particular the cosmological fine tuning problem  \cite{ArMuSt}.

One of the peculiar feature of the dark energy is a possibility of the phantom energy equation of state $\rho+p<0$. This phantom energy violates the weak energy dominance condition. In the case of phantom energy the cosmological scenario with the `Big Rip' is possible when cosmological phantom energy density grows at large times and disrupts finally all bounded objects up to subnuclear scale \cite{Caldw}. The other peculiarity is the diminishing of the black hole mass due to phantom energy accretion \cite{BDE}. The phantom energy is usually associated with the phantom or ghost fields: e.~g. the scalar fields with a wrong sign kinetic term \cite{Caldw,Cline}. The phantom energy has some peculiar properties in the framework of QFT in the curved space-time \cite{NojOdi}. The thermodynamic properties of the phantom energy is also rather unusual \cite{BreNojOdiVan}.

The existence of phantom energy is not excluded by the nowadays observations.  The measurements \cite{Ton03} of the distances and host extinctions of the 230 SN Ia provide the constraints on the dark energy equation of state, $-1.48<w<-0.72$.  In \cite{Alam} the data set containing 172 type Ia supernovas are analyzed in the model independent manner and it was shown that the presence of the
phantom energy with $-1.2<w<-1$ is preferable for the recent epoch.  The Chandra telescope observations \cite{All04} of the hot gas in the 26 X-ray luminous dynamically relaxed galaxy clusters provides $w=-1.20^{+0.24}_{-0.28}$, which is also in favor of the phantom energy.

The evolution of dark energy are considered usually in relation to
the cosmological problems. However the local evolution of self-gravitating dark energy may be quite different from the cosmological one. This is because of the nonlinearity of the General Relativity equations. The three dimensional analytical treatment is possible only in very restrictive cases, e.~g. in the case of the stationary accretion onto black hole of the dark energy considered as test fluid, i.~e. with a negligible self-gravitation \cite{BDE}.
self-gravitating fluid one must go to some simplified models.
One of the analytically treatable approach with a fluid self-gravitation is taking into account is  the thin shells model.  The theory of thin shells in General Relativity was developed by W.~Israel \cite{Isr66} and developed then by many authors (see e.~g. \cite{BerKuzTka} for review and references).  The problem of thin shell analysis is greatly simplified in the case of spherical symmetry. The aim of this paper is to consider several scenarios for the thin spherically symmetric phantom shell evolution.

The paper is organized as follows. In Sec.~\ref{GenTh} the general
concepts of spherically symmetric gravity are outlined with
special attention to Schwarzschild space-time. In Sec.~\ref{ThinSh}
the specific equations of motion for thin shells are obtained. In
Sec.~\ref{EvolPh} the evolution of shell with phantom equation of
state is analyzed and different types of motion are classified. In
Sec.~\ref{Dis} we briefly discuss the obtained results.
Throughout the paper we use the units $\hbar=c=1$.

\section{General Theory}
\label{GenTh}
\subsection{Spherically symmetric gravity}
\label{Ssg}

A spherically symmetric manifold is a direct product of a two-dimensional manifold $M_2$ and two-dimensional sphere $S_2$, that is, $M_4 = M_2 \times S_2$. The line element of any spherically symmetric space-time can always be written in the form
\begin{equation}
\label{ds}
ds^2 = g_{\alpha\beta} dx^{\alpha}dx^{\beta}\\
     = A dt^2 + 2H dt dq + B dq^2 + R^2 (t,q) d\Omega^2
\end{equation}
with the signature $(+,-,-,-)$. Here $t$ and $q$ are correspondingly the timelike and spacelike coordinates, $A$, $H$ and $B$ are functions of $t$ and $q$ only, and $R(t,q)$ is the radius of a two dimensional sphere (in the sense that the area of the sphere equals to $4\pi R^2)$,
\begin{equation}
d\Omega^2 = d\theta^2 + \sin \theta^2 d\phi^2
\end{equation}
being the line element of the unit sphere. For the given space-time the coefficients $A$, $H$ and $B$ are not uniquely defined. One can transform the line element (\ref{ds}) to the new coordinate system which conserves explicitly the spherically symmetric form of the metric:
\begin{equation}
\label{trans}
\tilde t = \tilde t(t,q), \quad
\tilde q = \tilde q(t,q).
\end{equation}
Unlike the metric coefficients in $M_2$, the radius $R(t,q,)$ is invariant under the transformation (\ref{trans}). The other very important invariant is constructed from the partial derivatives of $R$ as follows
\begin{equation}
\label{delta}
\Delta = \gamma^{\alpha \beta} R_{,\alpha} R_{,\beta},
\end{equation}
where $\gamma^{\alpha \beta}$ is inverse to the two-dimensional
metric tensor $\gamma_{\alpha \beta}$. This invariant is nothing
more but the square of the normal vector to the surface $R=const$.

If we know two invariant functions $R(t,q)$ and $\Delta (t,q)$ we know the line element of the spherically symmetric space-time up to the gauge transformations and, therefore, its local structure. To construct the global manifold we need some additional principle. Physics provides us with it. From the point of view of physicists any space-time should be geodesically complete \cite{MisThoWhe}, that is, every timelike and null geodesics should start and end either at infinities or at the singularities.

The function $\Delta (t,q)$ brings a nontrivial information about a space-time structure. Indeed, in the flat Minkowskian
space-time $\Delta \equiv -1$, all the surfaces $R = const$ are
timelike and therefore, $R$ can be chosen as spatial coordinate $q
= R$ on the whole manifold. But in the curved space-time $\Delta$
is no more constant and can in general be both positive and
negative. The region with $\Delta < 0$ is called the $R$-region,
and the radius can be chosen as a radial coordinate $q$. In the
region with $\Delta > 0$ the surfaces $R = const$ are spacelike
(the normal vector is timelike), and the radius $R$ can be chosen
as a time coordinate $t$. Such regions are called the $T$-regions.
The $R$- and $T$-regions were introduced by Igor Novikov. But this
is not the whole story. It is easy to show that we can not get
$\dot R = 0$ (``dot'' means a time derivative) in a $T$-region.
Hence it must be either $\dot R > 0$ (such a region of inevitable
expansion is called $T_+$-region) or $\dot R < 0$ (inevitable
contraction, a $T_-$-region). The same holds for $R$-regions. They
are divided in two classes, those with $R'>0$ (``prime'' stands
for a spatial derivative) which are called $R_+$-regions, and
$R_-$-regions with $R' < 0$. These, $R$- and $T$-regions are
separated by the surfaces $\Delta = 0$ which are called the
apparent horizons. The apparent horizons can be null, timelike or
spacelike.

Thus, the curved spherically symmetric space-times may in general have a rather complex structure, a set of $R_{\pm}$- and $T_{\pm}$-regions separated by apparent horizons $\Delta = 0$. In the next subsection we consider one of the important example of spherically symmetric manifolds.

\subsection{Schwarzschild space-time}
\label{Schw}

\begin{figure}[t]
\begin{center}
\includegraphics[angle=0,width=0.85\textwidth]{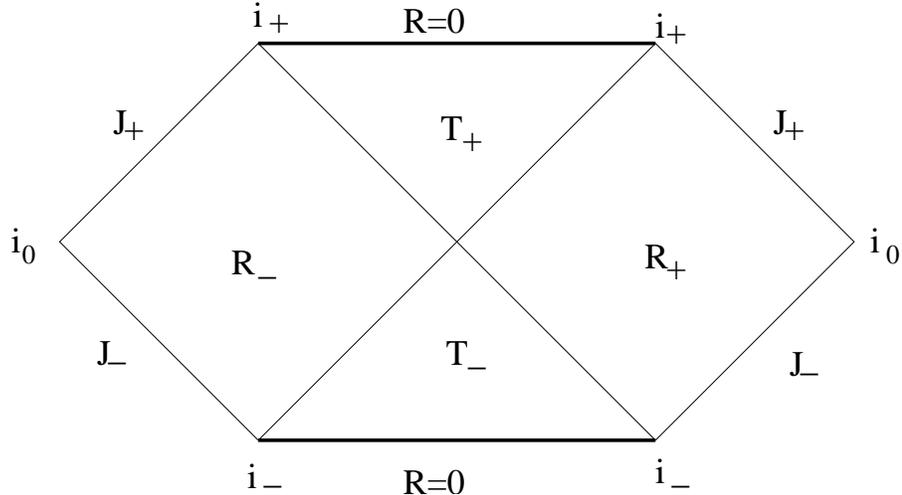}
\end{center}
\caption{Carter-Penrose diagram for Schwarzschild space-time. Every
point represents a sphere. Here $I_-$ and $I_+$ mean past and
future null infinities respectively. The past and future timelike and spacelike infinities are $i_-$ and $i_+$, $i_0$ respectively. Regions  $R$ and $T$ are separated by the two apparent horizons (the future
and past ones) which are null surfaces. In our case the future
horizon coincides with the event horizon defined as the first null
geodesics which does not reach infinity. Its time reversal is
called a particle horizon. The $T_-$-region is called the black
hole. The $T_+$-region is called the white hole, and the
$R_-$-region is called the wormhole respectively.}
\label{swrzld1}
\end{figure}

\begin{figure}[t]
\begin{center}
\includegraphics[angle=0,width=0.85\textwidth]{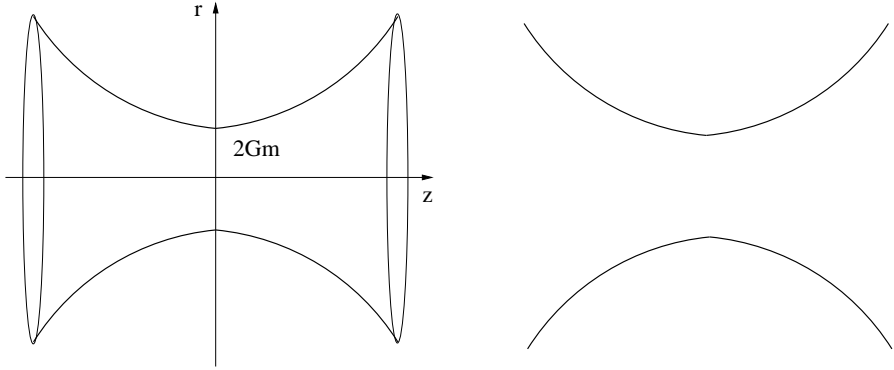}
\end{center}
\caption{Schwarzschild spacetime at fixed time and $\theta$ and
its embedding diagram. The throat width is proportional to
Schwarzschild mass.}\label{embed}
\end{figure}

The solutions to the vacuum Einstein equations consist of only one-parameter family. In the curvature coordinates ($q = R$) the
metric of a Schwarzschild space-time has the form
\begin{equation}
\label{schw}
ds^2 = F(R) dt^2 - F^{-1}(R) dR^2 - R^2 d\Omega^2,
\end{equation}
where
\begin{equation}
F(R) = 1 - \frac{2Gm}{R},\ \ \ m > 0
\end{equation}
and $G$ is the Newton gravitational constant. This static metric describes a space-time outside a spherically symmetric body with a mass $m$ called also a Schwarzschild mass. The metric (\ref{schw}) has a coordinate singularity at $R=2Gm$. It is related to the static nature of the line element (\ref{schw}) and impossibility to synchronize clocks of the observers who are static at the spatial infinity $(R = const\longrightarrow\infty)$ and those (who are nonstatic) in the region $R < 2Gm$. Moreover, it appears that the manifold described by the line interval (\ref{schw}) is not geodesically complete. It can be shown that the maximally extended Schwarzschild manifold consists of four parts. Furthermore, it is possible to choose such coordinates, which put infinities at finite distances. Schwarzschild geometry in such coordinates is represented by the Carter-Penrose diagram shown in the Fig.~\ref{swrzld1}.
An another useful representation of the Schwarzschild space-time is the so called embedding diagrams.
One can consider Schwarzschild metric with $t=\mbox{const}$ and $\theta= \pi/2$.
It is easy to show that this is the metric on a hyperboloid embedded in the three dimensional flat space
\begin{equation}
r=\frac{z^2}{8Gm} + 2Gm.
\end{equation}
So, an embedding looks like it is shown in the Fig.~\ref{embed}
In the following we will use schematic version of embedding  which
is called embedding diagram (right panel in the Fig.~\ref{embed}).
As one can see from the Fig.~\ref{embed} the wormhole is separated from the $R_+$-region (or the black hole exterior) by the throat called also the Einstein-Rosen bridge. Of course, the above definitions are by no means general, but they are sufficient for our purposes. (The general definitions require powerful mathematical tools and enormous number of predefinitions.) With the inclusion of matter sources the Schwarzschild solution is valid only outside their boundaries. In the case of the complete Schwarzschild manifold the sources are considered as concentrated at the past and future singularities at $R = 0$. The manifold is called the eternal black hole.

\section{Thin shells}
\label{ThinSh}

One of the most important feature of General Relativity is that the equations of motion of matter fields are incorporated into the Einstein equations. The Einstein equations of General Relativity are nonlinear partial differential equations. This means that the motion of test particles or fields on the given background will in general be different from that of the matter for the self-consistent solutions. It makes analysis very complicated. To obtain some definite results we have to choose the simplest possible model. This is a self-gravitating thin shell. In this section we derive equations of motion for thin shell.

Let us now introduce the notion of thin shell. Because we will be
dealing only with the timelike spherically symmetric thin shells we
adjust the very nice generally covariant formalism derived by W.~Israel \cite{Isr66} to the case of interest (see, e.g.
\cite{BerKuzTka}). Let us choose some hypersurface $\Sigma$
divided the whole space time into two parts, ``in'' and ``{\rm
out}''. With this hypersurface $\Sigma$ can be connected some
special coordinate system called Gauss normal coordinates. In our
simple case the line element written in these coordinates takes
the form
\begin{equation}
\label{normal}
ds^2 = d\tau^2 - dn^2 - R^2(\tau,n)d\Omega^2 .
\end{equation}
where $\tau$ is the proper time of the observer sitting fixed on
$\Sigma$, the coordinate $n$ grows from the ``in'' to the ``{\rm
out}''-region in the {\rm out}er normal direction to the
hypersurface $\Sigma$, $R(\tau,n)$ is the radius of the sphere
(in the sense that a sphere area equals to $4\pi R^2$) and $d\Omega^2$
is the line element of the unit sphere,
$d\Omega^2=d\theta^2+\sin^2\theta d\phi^2$. The hypersurface is
situated at $n=0$, and
\begin{equation}
\label{Sigma}
\begin{array}{rcl}
 ds^2_{\Sigma} &=& d\tau^2 -\rho^2(\tau) d\Omega^2,\\
\rho(\tau)&=&R(\tau,0)
\end{array}
\end{equation}
The hypersurface $\Sigma$ is called the singular shell if some energy momentum tensor is concentrated on it, namely $T_i^k=S_i^k\delta(n)+\dots$, $S_i^k$ is the surface energy-momentum tensor of the shell ($i,k$=0,2,3), otherwise the hypersurface is nonsingular. in our case due to the spherical symmetry the only nonzero components of $S_i^k$ are $S_0^0$ and $S_2^2=S_3^3$.

The introduced earlier invariant $\Delta$ equals
\begin{equation}
\label{normdelta}
\Delta=R,^2_\tau - R,^2_n
\end{equation}
and
\begin{equation}
\label{sigmadef}
R,_n|_\Sigma=\sigma \sqrt{ \dot\rho^2-\Delta},
\end{equation}
where $\sigma=+1$ if radii increase in the direction of the {\rm
out}er normal, and $\sigma=-1$ if radii decrease. Evidently,
$\sigma=+1$ in $R_+$-region and $\sigma=-1$ in $R_-$-region. Thus
on the equation of motion of the shell $\sigma$ can change sign
only in $T$-region, or in the region where its motion is
forbidden.

The subsequent procedure is very simple. Keeping in mind that the metric itself is continuous but some of its derivatives could undergo a jump across the shell, we integrate Einstein equations and obtain (the nontrivial result is only for the $(_0^0)$ and $(_2^2)$ components) after some algebra
\begin{equation}
\label{zerozero} \frac{2\sigma_{\rm
in}}{\rho}\sqrt{\dot\rho^2-\Delta_{\rm in}}- \frac{2\sigma_{\rm
out}}{\rho}\sqrt{\dot\rho^2-\Delta_{\rm out}}=8\pi G\rho S_0^0,
\end{equation}
\begin{equation}
\label{twotwo}
\begin{array}{rcl}
\frac{2\sigma_{\rm in}}{\rho}\sqrt{\dot\rho^2-\Delta_{\rm in}}-
\frac{2\sigma_{\rm out}}{\rho}\sqrt{\dot\rho^2-\Delta_{\rm out}}+
\frac{\sigma_{\rm in}}{\sqrt{\dot\rho^2 - \Delta_{\rm in}}}
\ddot\rho -
\frac{\sigma_{\rm out}}{\sqrt{\dot\rho^2 - \Delta_{\rm out}}} \ddot\rho + \\
\frac{\sigma_{\rm in}}{2\rho\sqrt{\dot\rho^2 - \Delta_{\rm
in}}}(1+\Delta_{\rm in})-
\frac{\sigma_{\rm out}}{2\rho\sqrt{\dot\rho^2 - \Delta_{\rm out}}}(1+\Delta_{\rm out})+\\
+4 \pi G \rho (^{({\rm out})}T_n^n -  ^{(in)}T_n^n)=8 \pi G S_2^2.
\end{array}
\end{equation}
The continuity equation for the energy-momentum tensor is
transformed to
\begin{equation}
\label{ssdot} \frac{dS_0^0}{d\tau} + \frac{2 \dot\rho}{\rho}
(S_0^0 - S_2^2) + ~^{({\rm out})}T_0^n - ~^{(in)}T_0^n = 0.
\end{equation}
The third equation is a differential consequence of the first two
ones. But very often it is convenient to use all three of them.

In what follows we will consider the thin shell in vacuum. So,
both inside and outside the shell we will have the Schwarzschild
metric with different masses (because the shell has its own mass
which is added to the inner mass) and our equations become:
\begin{equation}
\label{main1} \sigma_{\rm in}\sqrt{\dot\rho^2+F_{\rm
in}}-\sigma_{\rm out}\sqrt{\dot\rho^2+F_{\rm out}}=4\pi G\rho
S_0^0,
\end{equation}
\begin{equation}
\label{main2}
\ddot \rho = -4\pi^2 G^2 \rho (S_0^0)^2 + 8\pi^2G^2\rho S_0^0 S_2^2  \\
- \frac{G(m_{\rm in}+m_{\rm out})}{2\rho^2} - \frac{\Delta m^2
S_2^2}{8\pi^2\rho^5(S_0^0)^3}
\end{equation}
\begin{equation}
\label{main3} \dot S_0^0+\frac{2\dot\rho}{\rho}(S_0^0-S_2^2)=0,
\end{equation}
where $\Delta=-F=-1+\frac{2Gm}{\rho}$. We wrote the second equation in a somewhat different (twice squared) form, which is more suitable for us. The information about a global geometry (signs of $\sigma_{\rm in}$ and $\sigma_{\rm out}$) is already contained in the
equation (\ref{main1}).

\section{Evolution of phantom shell}
\label{EvolPh}

Now let us apply the above theory to phantom shells.
Consider a simple (but a rather general!) linear equation-of-state relation $S_0^0=kS_2^2$. In the phantom case $k>1$. The solution of (\ref{main3}) is
\begin{equation}
S_0^0=C\rho^{2(k-1)},
\end{equation}
where we call the constant $C$ the 'shell power'.  Denote also
\begin{equation}
\label{defx} x\equiv4\pi^2GC^2\rho^{4k-1}.
\end{equation}
After some manipulations with (\ref{main1}) and taking into account
(\ref{defx}) one gets the following two equations:
\begin{equation}
\label{dotrhox2} \dot\rho^2=-1+\frac{G}{\rho}\left((m_{\rm
in}+m_{\rm out})+\frac{\delta m^2}{4x}+x\right),
\end{equation}
\begin{equation}
\label{ddotrhox} \ddot\rho=-\frac{G}{2\rho^2}\left[(m_{\rm
in}+m_{\rm out})+\frac{k\delta m^2}{x}-2(2k-1)x\right]
\end{equation}
and the sign conditions
\begin{equation}
\label{signin} \sigma_{\rm in}={\rm sign}\left[\delta
m+8\pi^2G\rho^3(S_0^0)\right],
\end{equation}
\begin{equation}
\label{signout} \sigma_{\rm out}={\rm sign}\left[\delta
m-8\pi^2G\rho^3(S_0^0)\right]
\end{equation}
or in a more convenient form
\begin{equation}
\label{xsignin} \sigma_{\rm in}={\rm sign}\left(\delta
m+2x\right),
\end{equation}
\begin{equation}
\label{xsignout} \sigma_{\rm out}={\rm sign}\left(\delta
m-2x\right).
\end{equation}
Here, we denote $\delta m \equiv m_{\rm out}-m_{\rm in}$.

\begin{figure}[t]
\begin{tabular}[c]{cc}
\includegraphics[angle=0,width=0.45\textwidth]{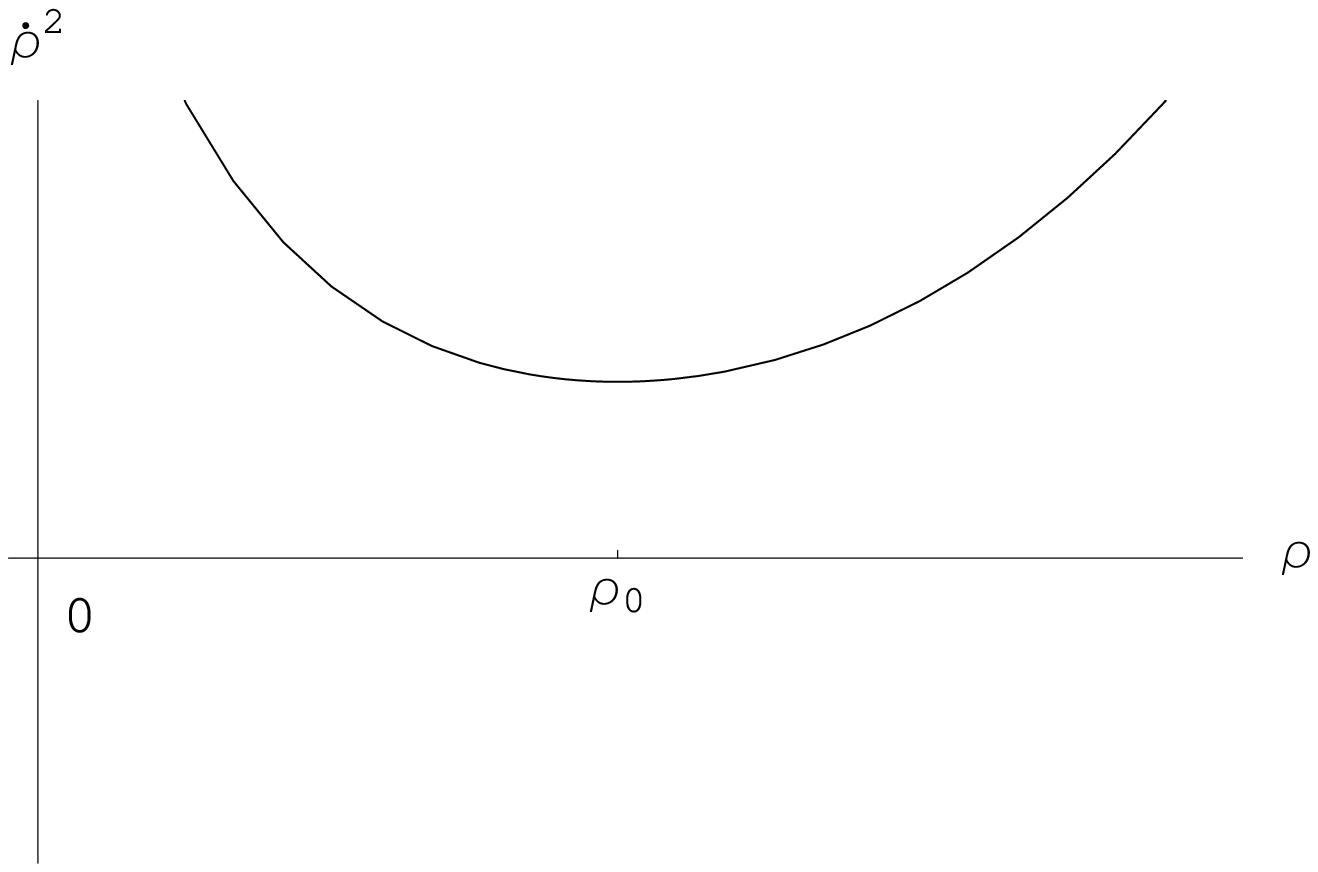} &
\includegraphics[angle=0,width=0.45\textwidth]{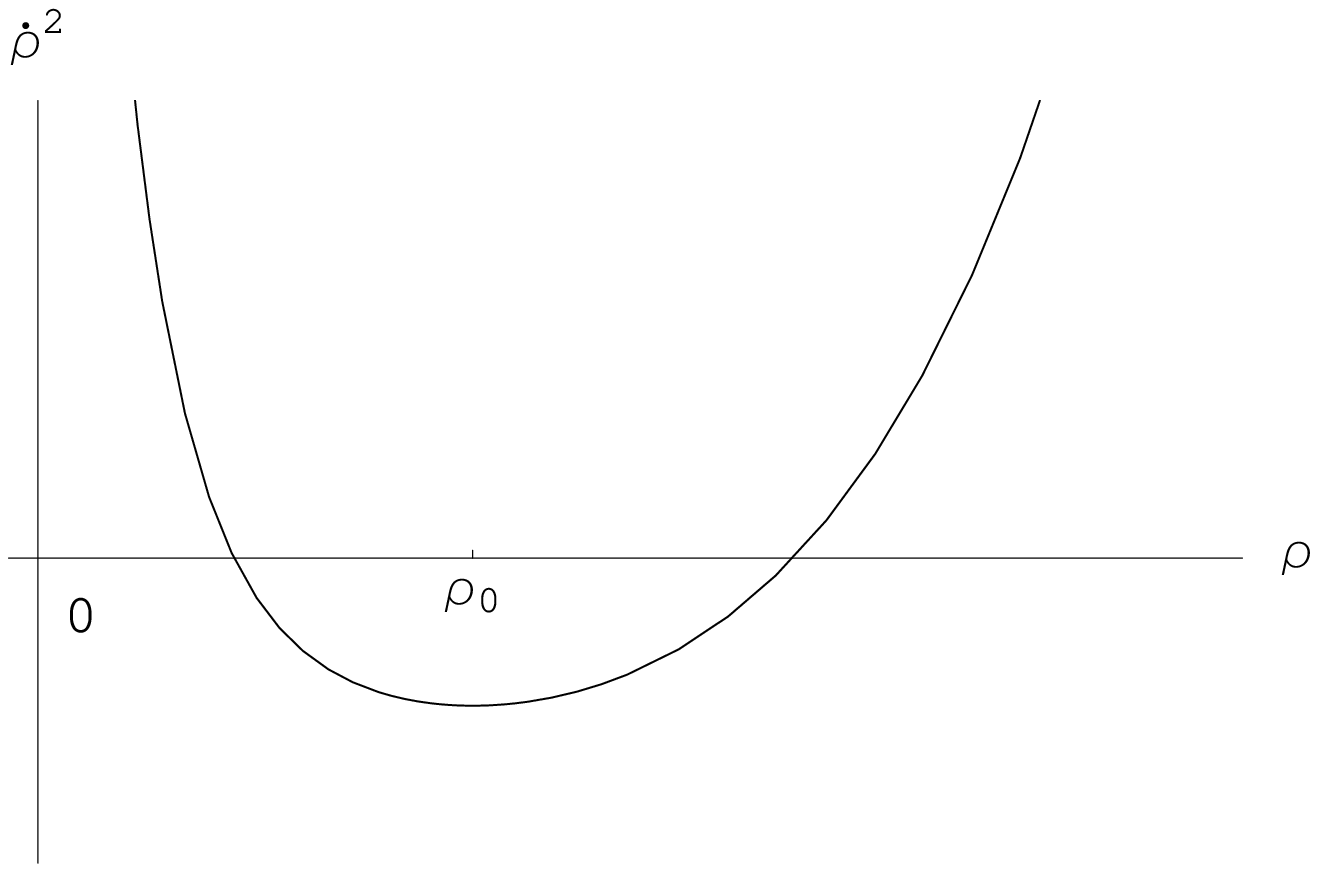} \\
\includegraphics[angle=0,width=0.45\textwidth]{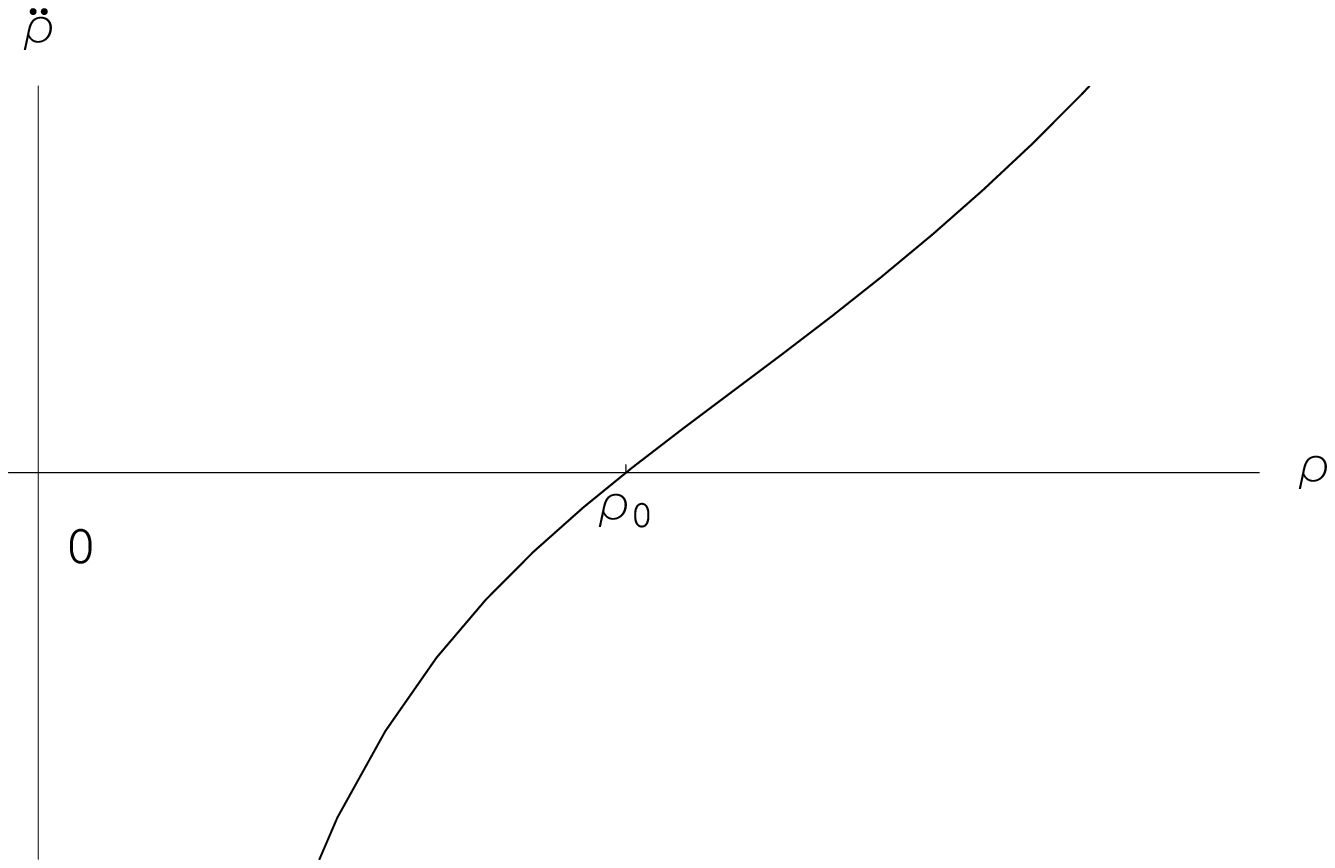} &
\includegraphics[angle=0,width=0.45\textwidth]{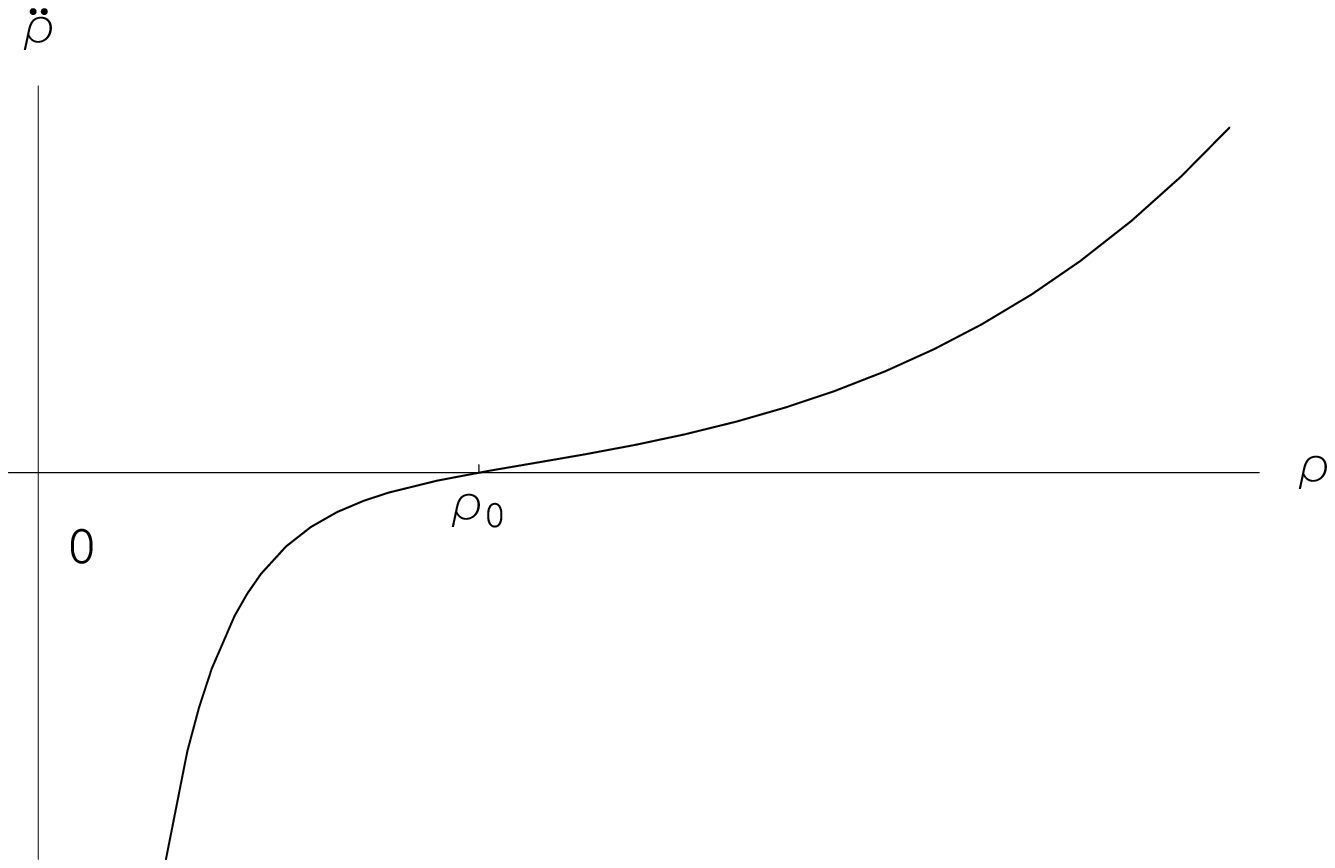}
\end{tabular}
\caption{
Schematic curves $\dot\rho^2(\rho)$ (up) and $\ddot\rho(\rho)$ (down) according to (\ref{dotrhox2}) and (\ref{ddotrhox}). The case
$C^2>C_0^2$ and $C^2<C_0^2$ corresponds to the left and right panel respectively. The region $\dot\rho^2<0$ is forbidden for motion.
}
\label{clec}
\end{figure}

The curves $\dot\rho^2(\rho)$ and $\ddot\rho(\rho)$ are shown in the Fig.~\ref{clec}. There are possible several evolution scenarios for the shell. When the curve $\dot\rho^2(\rho)$ lies completely above the $\rho$ axis, there is only the infinite motion. On the other case, when the function $\dot\rho^2(\rho)$ has roots, it is possible both the finite motion and infinite one.

As one can see from  equation (\ref{defx}), it is convenient to
define the parameter space of the problem using  $m_{\rm in}$, $m_{\rm
out}$, $k>1/4$ and $\rho$ as free parameters. Then, instead of  search for conditions imposed on $\rho$, it is enough to find conditions for the parameter $C$. This conditions will define the global geometry.
Now, let us construct these conditions. First of all, the change of  sign of the acceleration $\ddot\rho$ in (\ref{ddotrhox}) occurs when $\ddot\rho=0$. This corresponds to the quadratic equation, whose positive root is
\begin{equation}
\label{x0e} x_0=\frac{m_{\rm in}+m_{\rm out}+\sqrt{(4k-1)^2\delta
m^2+4m_{\rm in}m_{\rm out}}}{4(2k-1)}.
\end{equation}
The corresponding value of $\rho$ according to (\ref{defx}),
denoted by $\rho_0$. Consider the sign of $\dot\rho^2(\rho_0)$.
Let us introduce the parameter $C_0$ that $\dot\rho^2(\rho_0)>0$ when
\begin{equation}
\label{c0ltc}
C^2>C_0^2
\end{equation}
and $\dot\rho^2(\rho_0)<0$ when
\begin{equation}
\label{c0gtc}
 C^2<C_0^2
\end{equation}
The explicit value for $C_0$ is
\begin{eqnarray}
\!\!C_0&=&\left(\frac{x_0}{4\pi^2}\right)^{1/2}
\left[8k(2k-1\!)\right]^{\frac{4k-1}{2}}
\nonumber \\
&\!\times&\!\left\{ {(4k\!-\!1\!)G\!\left[(4k\!-\!1\!) (m_{\rm
in}\!+\!m_{\rm out})\!+\!\sqrt{(4k\!-\!1\!)^2\delta m^2\!+\!4m_{\rm
in}m_{\rm out}}\right]}\right\}^{\!\frac{1\!-\!4k}{2}}\!\!.
\label{c0}
\end{eqnarray}

\begin{figure}[t]
\begin{center}
\includegraphics[angle=0,width=0.95\textwidth]{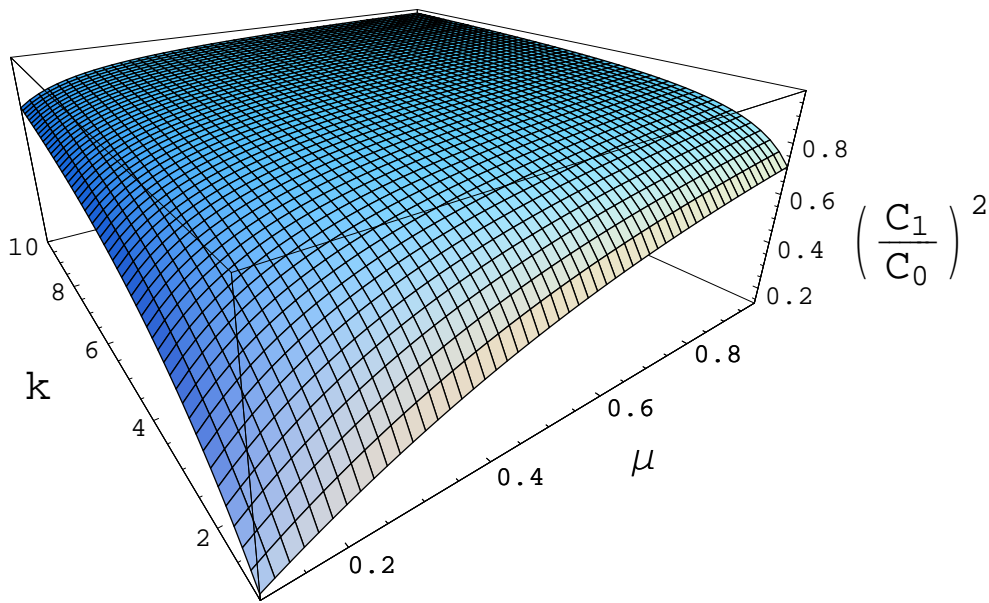}
\end{center}
\caption{Fraction $(C_1/C_0)^2$ as a function of $\mu$ and $k$.
\label{c1c0g} }
\end{figure}

Consider at first the case of $\delta m>0$. According to (\ref{xsignin}) there must be always $\sigma_{\rm in}=+1$. The $\sigma_{\rm out}$ changes sign at $x=x_1\equiv \delta m/2$ and $\sigma_{\rm out}=-1$ if $x>x_1$. Thus, in the case $x \to \infty$, one has $\sigma_{\rm out}=-1$, and, if $x \to 0$, then $\sigma_{\rm out}=+1$. The value of $\rho$ corresponding to $x_1$ is denoted by $\rho_1$. From (\ref{ddotrhox}) it is easy to see that
\begin{equation}
\ddot\rho(x_1)=-\frac{Gm_{\rm out}}{\rho_1^2}<0.
\end{equation}
From this follows the important conclusion that $\rho_1<\rho_0$.
This can also be proved by the direct comparison of $x_0$ and
$x_1$. From (\ref{dotrhox2}) one obtains
\begin{equation}
\dot\rho^2(x_1)=-1+\frac{2Gm_{\rm out}}{\rho_1}.
\end{equation}
Let us denote by $C_1$ such a value of $C$ that $\rho_1=2Gm_{\rm
out}$ at $C=C_1$:
\begin{equation}
C_1^2=\frac{\delta m}{2^{2(2k+1)}\pi^2m_{\rm out}^{4k-1}}.
\label{c1}
\end{equation}
In the case $\rho_1>2Gm_{\rm out}$ one has
 \begin{equation}
\label{c1gtc}
C^2<C_1^2
\end{equation}
and $\sigma_{\rm out}$ changes sign in the forbidden for motion
part of $R$-region. And vice versa, if $\rho_1<2Gm_{\rm out}$, then
\begin{equation}
\label{c1ltc}
C^2>C_1^2
\end{equation}
and $\sigma_{\rm out}$ changes sign in $T$-region. Now we have to define which of the conditions $C_0^2>C_1^2$ or
$C_1^2<C_0^2$ is true. This can be done by considering limits for
the parameters in (\ref{c0}) and(\ref{c1}) or by direct numerical
calculation of the fraction $(C_1/C_0)^2$ as a function of
$\mu=\delta m/m_{\rm out}$ and $k$. The results of such
calculation is shown in the Fig.~\ref{c1c0g}. It is clear now that
$C_1^2<C_0^2$.

It is obvious from the above analysis, that there exist three possible scenarios for the shell evolution in the case $\delta m>0$. In short, these scenarios are the following:
\begin{enumerate}
  \item  Infinite motion.  The sign of $\sigma_{\rm out}$ changes in the $T$ region. The shell power obeys the following inequality:
\begin{equation}
C^2>C_0^2.
\end{equation}
The embedding diagram for this scenario is shown in the Fig.~\ref{ediag1}.
\begin{figure}
\begin{center}
\includegraphics[angle=0,width=0.7\textwidth]{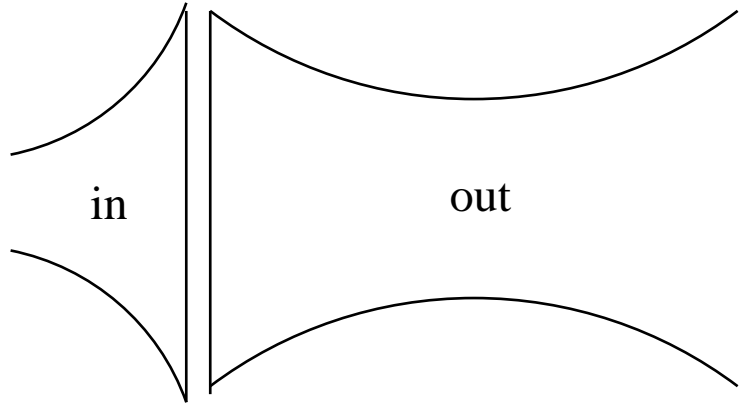}
\end{center}
\caption{Infinite inflation or collapse (symmetric in time). The
left throat is the more narrow then the right one because $\delta
m>0$.} \label{ediag1}
\end{figure}
The Carter-Penrose diagrams for the case of collapse and inflation are the same up to the
time reverse (see Fig.~\ref{diag1}).
\begin{figure}[t]
\begin{center}
\includegraphics[angle=0,width=0.95\textwidth]{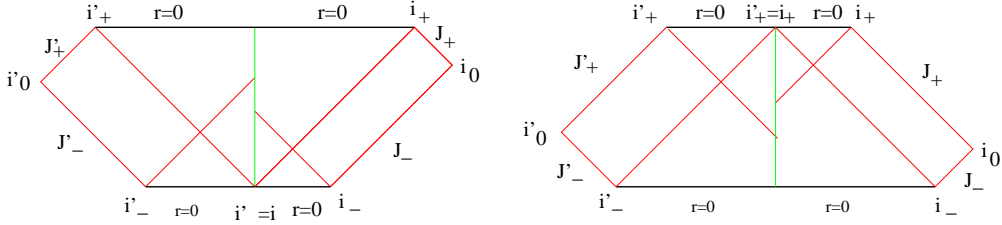}
\end{center}
\caption{The Carter-Penrose diagrams for a collapsing (left)
and inflating (right) shell.} \label{diag1}
\end{figure}
  \item There exist the turn points and $\sigma_{\rm out}$ changes sign in $T$-region:
 \begin{equation}
 C_1^2<C^2<C_0^2.
 \end{equation}
In this case the embedding diagrams are the same as in the previous case.
The corresponding Carter-Penrose diagrams are shown in the Fig.~\ref{diag2}
\begin{figure}[t]
\begin{center}
\includegraphics[angle=0,width=0.9\textwidth]{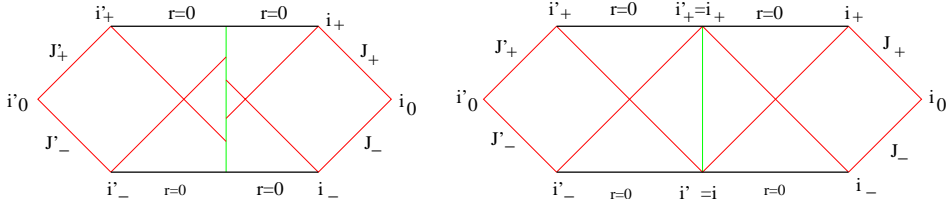}
\end{center}
\caption{The Carter-Penrose diagrams for a collapsing shell (left) and for a shell moving from infinity to infinity (right).} \label{diag2}
\end{figure}

  \item There exist the turn points and $\sigma_{\rm out}$ changes the sign in a part of $R$-region, which is forbidden for the motion.
  \begin{equation}
 C^2<C_1^2
 \end{equation}
In this case we have two embedding diagrams (see the Fig.~\ref{ediag12}).
\begin{figure}[t]
\begin{center}
\includegraphics[angle=0,width=0.95\textwidth]{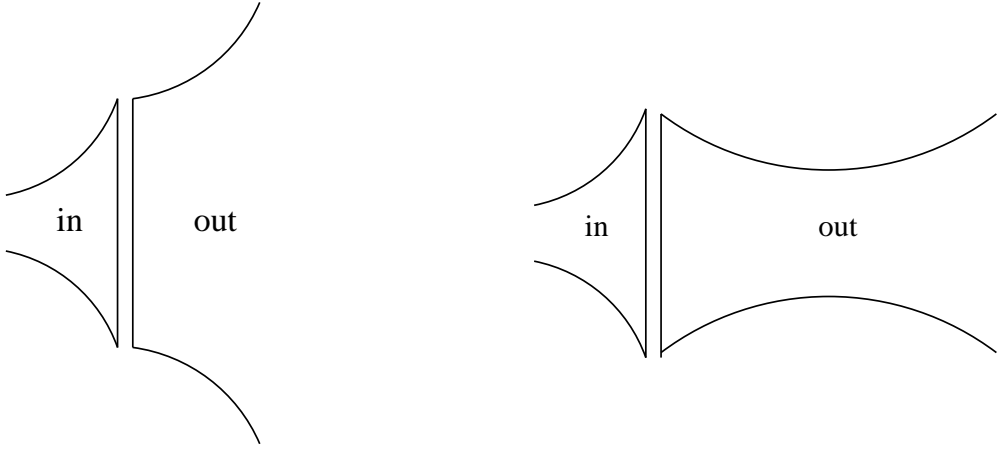}
\end{center}
\caption{The left diagram represents the case when the shell is at the left from the left turn point. In this case the shell is collapsed finally. At the right diagram the shell goes from the past infinity to the future infinity when it evolves toward the right from the right turn point.} \label{ediag12}
\end{figure}
The Carter-Penrose diagrams  are shown in the Fig.~\ref{diag3}
\begin{figure}[t]
\begin{center}
\includegraphics[angle=0,width=0.99\textwidth]{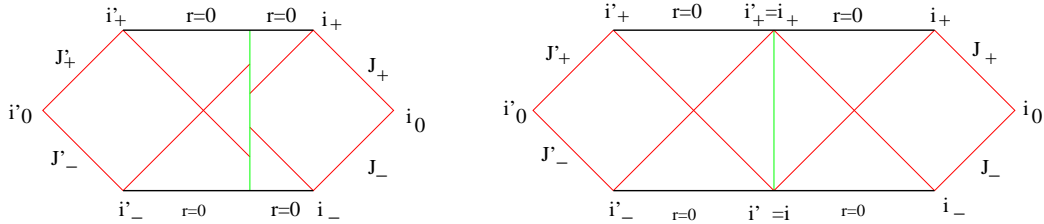}
\end{center}
\caption{The Carter-Penrose diagrams for the collapsing shell (left) and for the shell moving from infinity to infinity (right). In the both cases the shell evolves in $T_\pm$-  and $R_-$-regions.} \label{diag3}
\end{figure}
As follows from the above analysis if $\delta m>0$, the right
diagram in the Fig.~\ref{diag3} is the only case when the shell shows
itself in the $R_+$-region.
\end{enumerate}
Consider now the $\delta m<0$ case. The 'density' $S_0^0$ is assumed
to be always positive. So the negativity of $\delta m$ is caused exclusively by the gravitational mass defect. The $\sigma_{\rm in}$ changes sign at $x=x_2\equiv -\delta m/(2)$, and $\sigma_{\rm in}=+1$ if $x>x_2$. At the same time $\sigma_{\rm out}=-1$ Thus, we can conclude that in the case $x \to \infty$, one has $\sigma_{\rm in}=+1$ and $\sigma_{\rm in}=-1$ if $x \to 0$. Denote according to (\ref{defx}) the corresponding to $x_2$ value of $\rho$ by $\rho_2$. From (\ref{ddotrhox}) it is easy to see that
\begin{equation}
\ddot\rho(x_2)=-\frac{Gm_{\rm in}}{\rho_2^2}<0.
\end{equation}
From (\ref{dotrhox2}) one obtains
\begin{equation}
\dot\rho^2(x_2)=-1+\frac{2Gm_{\rm in}}{\rho_2}.
\end{equation}
Let us denote also by $C_2$ such a value of $C$ that $\rho_2=2Gm_{\rm
in}$ at $C=C_2$:
\begin{equation}
C_2^2=\frac{\delta m}{2^{2(2k+1)}\pi^2m_{\rm out}^{4k-1}}.
\end{equation}
In the case $\rho_2>2Gm_{\rm in}$ one has
\begin{equation}
C^2<C_2^2
\label{c2gtc}
\end{equation}
and the considered region is $R$-region.  And vice verse, if  $\rho_2<2Gm_{\rm in}$ then
\begin{equation}
C^2>C_2^2
\label{c2ltc}
\end{equation}
and one has $T$-region here. Just as in the case $\delta m >0$, the fraction $(C_2/C_0)^2 < 1$ because the 2d-graph is analog of the Fig.~\ref{c1c0g} with the replacement $\mu \to -\mu$. As a rsult, we can conclude that in the case  $\delta m <0$ there exist two evolution scenarios:
\begin{enumerate}
  \item Infinite motion.  The sign of $\sigma_{\rm in}$ changes in
$T$ region. The inequality for $C$-parameters are
\begin{equation}
C^2>C_0^2.
\end{equation}
The embedding diagram for this scenario is shown in the
Fig.~\ref{ediag100}.
\begin{figure}
\begin{center}
\includegraphics[angle=0,width=0.7\textwidth]{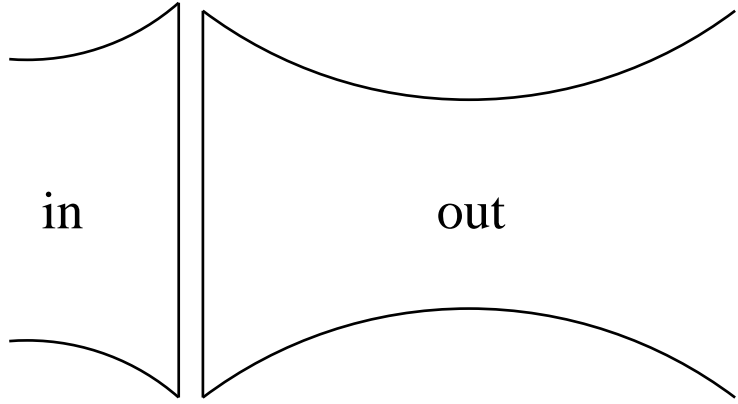}
\end{center}
\caption{Infinite inflation or collapse (symmetric in time). The
right throat is the more narrow then the left one because $\delta
m>0$.} \label{ediag100}
\end{figure}
The Carter-Penrose diagrams for the case of the collapse and
inflation are the same up to the time reverse. (see the
Fig.~\ref{diag5}).
\begin{figure}[t]
\begin{center}
\includegraphics[angle=0,width=0.95\textwidth]{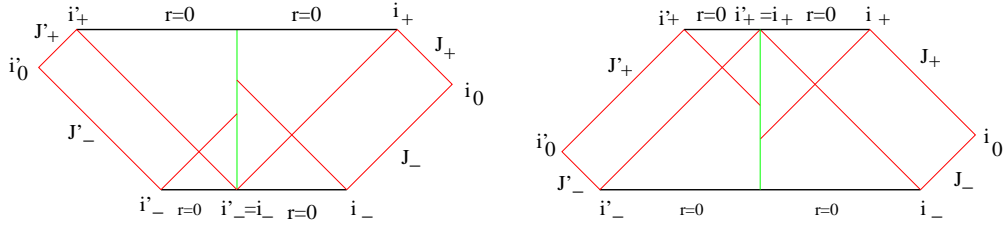}
\end{center}
\caption{The left Carter-Penrose diagram represents a collapsing
shell. The right one represents an inflating shell.} \label{diag5}
\end{figure}

  \item There exist the turn points and $\sigma_{\rm out}$ changes sign in $T$-region.
 \begin{equation}
 C_2^2<C^2<C_0^2.
 \end{equation}
In this case the embedding diagrams are the same as in the previous case. The corresponding Carter-Penrose diagrams are shown in the Fig.~\ref{diag6}
\begin{figure}[t]
\begin{center}
\includegraphics[angle=0,width=0.95\textwidth]{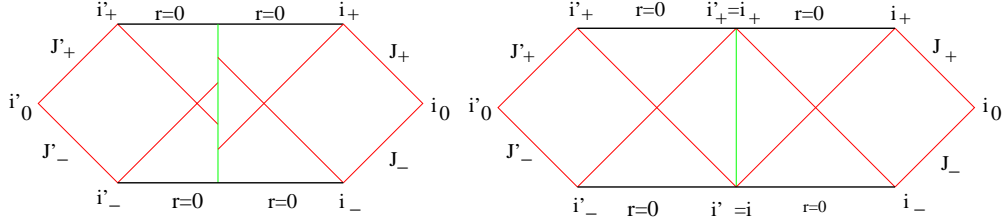}
\end{center}
\caption{The left Carter-Penrose diagram represents collapse of
the  shell. At the right one the shell  moves from infinity to
infinity. In both cases shell evolves in $T_\pm$-  and
$R_-$-regions (for a distant observer). } \label{diag6}
\end{figure}

  \item There exist the turn points and $\sigma_{\rm out}$ changes sign in a part of $R$-region, which is forbidden for the motion.
  \begin{equation}
  C^2<C_2^2.
  \end{equation}
For this case we have two embedding diagrams (see the Fig.~\ref{ediag34}).
\begin{figure}[t]
\begin{center}
\includegraphics[angle=0,width=0.95\textwidth]{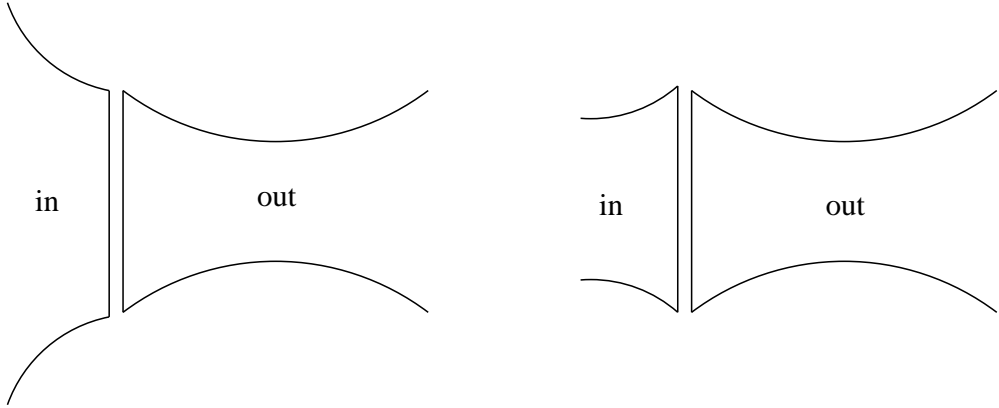}
\end{center}
\caption{The left diagram represents situation when the shell is
on the left from the left turn point. In this case the shell is
collapsed. At the right diagram shell goes from past  infinity to
future infinity  when it evolves to the right from the right turn
point.} \label{ediag34}
\end{figure}
The Carter-Penrose diagrams  are shown in the Fig.~\ref{diag78}.
\begin{figure}[t]
\begin{center}
\includegraphics[angle=0,width=0.95\textwidth]{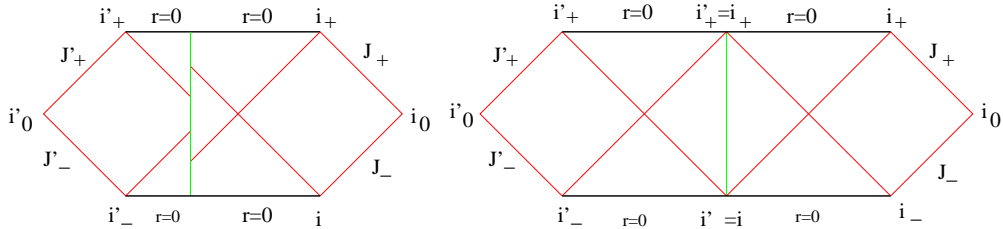}
\end{center}
\caption{The left Carter-Penrose diagram represents collapse of
the  shell. At the right one shell  moves from infinity to
infinity. In both cases shell evolves in $T_\pm$-  and
$R_-$-regions} \label{diag78}
\end{figure}
\end{enumerate}
We see again that for the all scenarios in the case of $\delta m <0$ the shell evolves under horizons and cannot reach a distant observer living in the $R_+$-region.

\section{Conclusion}
\label{Dis}

Discussion

We considered a dynamics of phantom thin shell surrounded
Schwarzschild black hole. The motivation for this work is the fact
that in many physically interesting situations in cosmology and
astrophysics the essential role was played the full account for
gravitational backreaction. In our case of phantom shells such a
backreaction may appear crucial for formation of the global
geometry of the space-time. The matter is that in General
Relativity any type of energy is gravitating. That is, not only
energy density but also the tension and pressure are gravitating.
The pressure plays a twofold role. The positive pressure causes
both repulsion and attraction, the later is due to its
contribution to the gravitating source. The negative pressure, on
the contrary, leads to the gravitational repulsion (the famous
example is the deSitter space-time). Hence the phantom shell is
even more repulsive. And indeed, we show that the global geometry
of the system consisting of the Schwarzschild surrounded by the
phantom shell is the wormhole-like type in all but one cases. In
the wormhole-like type geometry the distant observer cannot see
the shell at all, they are separated by the throat (Einstein-Rosen
bridge). The only exception is the case of the bound motion with
$\delta m>0$. But, though the distant observer may see the shell
it can not register the energy flux of the shell.

We are sure that despite the very simple character of our model
the result obtained should be taken into account in doing
calculation in cosmology and astrophysics when phantom energy is
present.

This work was supported in part by the Russian Foundation for
Basic Research grants 02-02-16762-a, 03-02-16436-a and
04-02-16757-a and the Russian Ministry of Science grants
1782.2003.2 and 2063.2003.2.

\end{document}